\documentclass[%
reprint,
showpacs,
bibnotes,
 amsmath,amssymb,
 aps,
pra]{revtex4-1}
\usepackage{graphicx}
\usepackage{bm}

\begin{document}

\title{Radiative correction in approximate treatments of electromagnetic scattering by point
and body scatterers}

\author{Eric C. Le Ru} \email{eric.leru@vuw.ac.nz}
\author{Walter R. C. Somerville}
\author{Baptiste Augui\'e}

\affiliation{The MacDiarmid Institute for Advanced Materials and Nanotechnology,
School of Chemical and Physical Sciences, Victoria University of Wellington,
PO Box 600, Wellington 6140, New Zealand}

\date{\today}

\begin{abstract}
The transition-matrix ($T$-matrix) approach provides a general formalism to study scattering problems in various areas of physics, including
acoustics (scalar fields) and electromagnetics (vector fields), and is related to the theory of the scattering matrix ($S$-matrix) used in quantum mechanics and quantum field theory.
Focusing on electromagnetic scattering, we highlight an alternative formulation of the $T$-matrix approach,
based on the use of the reactance matrix or $K$-matrix,
which is more suited to formal studies of energy conservation constraints (such as the optical theorem).
We show in particular that electrostatics or quasi-static approximations can be corrected within this framework
to satisfy the energy conservation constraints associated with radiation.
A general formula for such a radiative correction is explicitly obtained, and empirical expressions proposed in earlier studies
are shown to be special cases of this general formula.
This work therefore provides a justification of the empirical 
radiative correction to the dipolar polarizability and a generalization of this correction to any types of point or body scatterers of arbitrary shapes,
including higher multipolar orders.
\end{abstract}

\maketitle

\section{Introduction}

Radiative reaction, also known as radiation damping, refers to the fact that the electromagnetic (EM) field created by a charge or emitter must affect its own dynamics (e.g.~motion or power radiated) \cite{1954Heitler}. When applied to elementary charges \cite{1941Heitler}, no satisfactory classical treatment
of this effect has been found \cite{1998Jackson}, yet the radiative reaction is at the core
of the concepts of self-energy and renormalization in quantum electrodynamics (QED) \cite{1954Heitler}. In fact, using a Green's function approach
and regularization techniques akin to those of QED, a classical treatment of the radiative reaction for point electric dipole scatterers
can be obtained \cite{1998deVriesRMP}.

Interestingly, an equivalent result had been obtained
heuristically by adding a reaction field postulated from simple energy conservation arguments \cite{1982WokaunPRL}.
These arguments were inspired by research into simple models of the optical properties of sub-wavelength particles, notably for applications
in plasmonics and surface-enhanced Raman spectroscopy \cite{2009Book}, and we briefly present a similar derivation here.
The main idea is to use the solution of the electrostatics
problem to derive an approximate dipolar polarizability $\alpha_0$ for the particle. $\alpha_0$, assumed isotropic here
for simplicity, defines the electrostatics response of the particle and is such that a uniform external electrostatic field
$\mathbf{E}_0$ induces a dipole moment $\mathbf{p}_0=\alpha_0\mathbf{E}_0$.
In such an electrostatics problem, the power absorbed by the particle equals the work done by the external field on the charges \cite{1998Jackson}
and is therefore $P^0_\mathrm{abs}=(1/2)\omega\mathrm{Im}(\alpha_0)|\mathbf{E}_0|^2$.

In the {\it electrostatics, or quasi-static, approximation}, also often called Rayleigh approximation (see e.g. Chap. 5 in Ref. \onlinecite{1983Bohren}),
the far-field optical response of a sub-wavelength scatterer to an incident electric field $\mathbf{E}_\mathrm{inc}$
oscillating at frequency $\omega$ is approximated as that of an oscillating induced dipole given by $\mathbf{p}=\alpha\mathbf{E}_\mathrm{inc}$.
Note that we use complex notations with the $\exp(-i\omega t)$ convention and also that SI units are used throughout.
We also define the wave-vector in the medium $k_1=\sqrt{\epsilon_1}\omega/c$.
The electrostatics approximation consists in approximating $\alpha$ by $\alpha_0$, even if the new polarizability
$\alpha$ is in principle different from $\alpha_0$ because the electrostatics solution does not account for radiation.
The power radiated (or scattered) in this approximation is therefore $P_\mathrm{sca}=(\omega k_1^3 |\alpha_0|^2 |\mathbf{E}_\mathrm{inc}|^2)/(12\pi\epsilon_0 \epsilon_1)$. The power absorbed by the particle is approximated by its electrostatics value $P_\mathrm{abs}=P^0_\mathrm{abs}$.
The extinguished power is the power extracted by this point dipole from the incident EM field (the work of the field on the dipole) and is $P_\mathrm{ext}=(1/2)\omega\mathrm{Im}(\alpha)|\mathbf{E}_\mathrm{inc}|^2$ in the general case. It therefore reduces in the electrostatics approximation  ($\alpha \approx \alpha_0$) to $P_\mathrm{ext}=P_\mathrm{abs}$, which appears to contradict the
energy conservation condition $P_\mathrm{ext}=P_\mathrm{sca}+P_\mathrm{abs}$. This is not so surprising since
the electrostatics solution does not account for radiation (scattering) effects.
In fact, there is no contradiction in the strict range of validity of the electrostatics approximation, i.e. in the limit of vanishing size, as
we then have $P_\mathrm{sca} \ll P_\mathrm{ext},P_\mathrm{abs}$ since $\alpha_0$ scales with particle volume.
Nevertheless, it is useful in many instances to correct this problem to extend the range of applicability of the electrostatics approximation.
This can be achieved, as proposed in Ref. \onlinecite{1982WokaunPRL}, by defining a radiation-corrected polarizability, which
by construction enforces the energy conservation condition $P_\mathrm{ext}=P_\mathrm{sca}+P_\mathrm{abs}$, i.e.:
\begin{equation}
\mathrm{Im}(-\frac{1}{\alpha^\mathrm{RC}})=\frac{\mathrm{Im}(\alpha^\mathrm{RC})}{|\alpha^\mathrm{RC}|^2}=\mathrm{Im}(-\frac{1}{\alpha_0})+\frac{k_1^3}{6\pi\epsilon_0
\epsilon_1}.
\end{equation}
This condition on the imaginary part only is not sufficient to define $\alpha^\mathrm{RC}$ uniquely (unless Kramers-Kr\"onig relations \cite{1956TollPR} are
used) and the additional condition that
$\mathrm{Re}({1}/{\alpha^\mathrm{RC}})=\mathrm{Re}({1}/{\alpha_0})$ is usually assumed without further justification to obtain the radiative correction
to the polarizability as:
\begin{equation}
\frac{1}{\alpha^\mathrm{RC}}=\frac{1}{\alpha_0}-i\frac{k_1^3}{6\pi\epsilon_0 \epsilon_1},
\label{EqnRadCor}
\end{equation}
or equivalently \cite{1982WokaunPRL}
\begin{equation}
\alpha^\mathrm{RC}=\frac{\alpha_0}{1-i\frac{k_1^3}{6\pi\epsilon_0
\epsilon_1}\alpha_0}.
\end{equation}
This expression can also be derived rigorously in the special case of spherical particles by expansion of the Mie coefficients \cite{1983MeierOL,1983Bohren}
and can be generalized to spheroidal particles \cite{1982WokaunPRL,2003KellyJPCB}.
This corrected polarizability has been used in numerous contexts,
including for example surface-enhanced Raman scattering \cite{1982WokaunPRL,2009Book}, plasmonics \cite{1997NovotnyJAP,2003KellyJPCB,2004FordPRB},
or the discrete dipole approximation \cite{1992Lakhtakia,1994Draine,2006Novotny,2007YurkinJQSRT}.
More recently, this radiative correction has also been generalized, using again a heuristic approach based
on the optical theorem, to the case of point magnetic dipole and
dipolar scatterers with magneto-electric coupling \cite{2011SersicPRB}, for the study of metamaterials.

In this paper, we propose a general framework for the study and further understanding of the concept of radiative correction,
based on a simple reformulation of the $T$-matrix approach to EM scattering \cite{1966Newton,2002Mishchenko}.
The $T$-matrix formalism, often used in conjunction with the Extended Boundary Condition Method (EBCM), or null-field method,
was introduced more than 40 years ago \cite{1965Waterman} and is arguably one of the most
elegant and efficient methods to solve problems of electromagnetic scattering by particles of arbitrary shape and size \cite{1971WatermanPRD,1975BarberAO,1990Barber,2000Tsang,2002Mishchenko,2006Doicu,2006Martin}. It has been
applied, for example, to the study of scattering
by aerosols \cite{2007Yang}, metallic nanoparticles \cite{2009BoyackPCCP,2007KhlebtsovJPCC,1983BarberPRL,1983BarberPRB},
and collections of spheres \cite{1996MackowskiJOSAA}, and
also to more formal studies of EM scattering \cite{2010XuPRA}.
It has also been used extensively for acoustic scattering \cite{1969WatermanJASA,2006Martin}.

Our reformulation emphasizes the important role of the {\it reactance matrix}, or $K$-matrix \cite{1966Newton,1972Taylor}
in relation to energy conservation and radiative correction.
Although the $K$-matrix has been used occasionally in the past
in the context of the quantum theory of scattering \cite{1966Newton,1967TobockmanPR,1968LandshoffJMP,1972Taylor}, it seldom
appears in EM theory.
We show that all the aforementioned results for the radiative correction in EM scattering are special cases
of a general formula derived in this work. In addition to highlighting the importance of the $K$-matrix for general scattering theory,
this work therefore provides a formal justification of
existing radiative correction formulae and a generalization applicable to any type of point scatterer or particle
of arbitrary shape.
The latter point is a direct consequence of the fact that the $T$-matrix formulation of EM scattering is extremely general.
It applies to particles of arbitrary shape and may also cover for example \cite{2002Mishchenko,2006Doicu}
optically-active or anisotropic materials, layered particles, and multiple scattering by collections of particles.
The proposed $K$-matrix reformulation and associated radiative correction are therefore applicable to all the aforementioned cases.

The paper is organized as follows: in Sec.~\ref{SecTmatrix}, we briefly review the general principles of
the $T$-matrix approach to EM scattering. We then introduce in Sec.~\ref{SecKmatrix} an alternative, but closely related,
 formulation of the problem in terms of the $K$-matrix. In Sec.~\ref{SecRadCor}, we discuss the implications of
 the $K$-matrix formulation with regard to radiative corrections and obtain a general formula (Eq.~\ref{EqnRCESA}) for
 the radiative correction in EM scattering. Finally, in Sec.~\ref{SecAppl}, we show explicitly how this formula applies
 to specific cases of radiative correction that have been presented in the literature, therefore justifying, and
 in some cases extending, these previously empirical results.

\section{The $T$-matrix approach}
\label{SecTmatrix}

\subsection{Definition of the $T$-matrix}

We consider the general problem of electromagnetic scattering by a body characterized by a
linear local isotropic relative dielectric function $\epsilon_2$ (possibly frequency-dependent) 
embedded in a non-absorbing medium of refractive index $n_1$ (and relative dielectric function $\epsilon_1=n_1^2$).

Within the $T$-matrix approach \cite{1971WatermanPRD,1975BarberAO},
the EM field solution is expanded in a basis of vector spherical wave functions (VSWFs) in a similar fashion as for Mie theory 
\cite{1983Bohren}. We here follow the conventions of Mishchenko \cite{2002Mishchenko} for the definition of the VSWFs (see Appendix \ref{AppVSWF} for details).
The incident field $\mathbf{E}_\mathrm{inc}$ and internal field $\mathbf{E}_\mathrm{int}$ (the field in the region inside the particle) are regular
at $r=0$ and can therefore be expressed in terms of regular VSWFs denoted $\mathbf{M}^{(1)}_{\nu},\mathbf{N}^{(1)}_{\nu}$.
The scattered field $\mathbf{E}_\mathrm{sca}$ must satisfy the Sommerfeld radiation condition
and is therefore expanded in terms of outgoing spherical waves VSWFs denoted $\mathbf{M}^{(3)}_{\nu},\mathbf{N}^{(3)}_{\nu}$.
Explicitly:
\begin{eqnarray}
	\mathbf{E}_\mathrm{inc}(\mathbf{r}) &= \sum_{\nu}{a_{\nu}\mathbf{M}^{(1)}_{\nu}(k_1\mathbf{r}) + b_{\nu}\mathbf{N}^{(1)}_{\nu}(k_1\mathbf{r})},\nonumber\\
	\mathbf{E}_\mathrm{sca}(\mathbf{r}) &= \sum_{\nu}{p_{\nu}\mathbf{M}^{(3)}_{\nu}(k_1\mathbf{r}) + q_{\nu}\mathbf{N}^{(3)}_{\nu}(k_1\mathbf{r})},\nonumber\\
	\mathbf{E}_\mathrm{int}(\mathbf{r}) &= \sum_{\nu}{c_{\nu}\mathbf{M}^{(1)}_{\nu}(k_2\mathbf{r}) + d_{\nu}\mathbf{N}^{(1)}_{\nu}(k_2\mathbf{r})},
\label{EqnTexp}
\end{eqnarray}
where $k_i=(2\pi/\lambda)\sqrt{\epsilon_i}$ ($i=1,2$) are the wavevector amplitudes in region 1 (outside) and 2 (inside)
and $\lambda$ is the excitation wavelength.
These expansions can be represented as vectors, for example 
$(p_\nu,q_\nu)\equiv (\mathbf{p},\mathbf{q})$ for the scattered field, where the index $\nu=(n,m)$ combines the total ($n$) and projected ($|m|\le n$) angular momentum indices.
The expansion of the incident field $(a_\nu,b_\nu)$ for a given scattering problem is known, with explicit expressions
existing for example for plane waves \cite{2002Mishchenko}.

By linearity of Maxwell's equations, the coefficients of the scattered field are linearly related to those
of the incident field. This can be expressed explicitly by introducing the $T$-matrix:
\begin{equation}
\begin{pmatrix} \mathbf{p}\\ \mathbf{q}\end{pmatrix}=\mathbf{T}
\begin{pmatrix} \mathbf{a}\\ \mathbf{b}\end{pmatrix},
\label{EqnTcoeff}
\end{equation}
where $\mathbf{T}$ is an infinite square matrix,
which can be written in block notation as:
\begin{equation}
\mathbf{T}=
\begin{pmatrix}
\mathbf{T}^{11}& \mathbf{T}^{12} \\[0.4cm]
\mathbf{T}^{21}&\mathbf{T}^{22}
\end{pmatrix}.
\label{EqTmatBlock}
\end{equation}

In principle, from a knowledge of the $T$-matrix (at a given wavelength), one can infer the scattering properties
for any incident excitation. The $T$-matrix approach is therefore particularly suited for
computations of the scattering properties of a collection of randomly oriented scatterers \cite{1991MishchenkoJOSAA},
which is indeed one of the important applications of this formalism \cite{2002Mishchenko}.

We note that linear relationships involving the expansion coefficients
of the internal field can also be written:
\begin{equation}
\begin{pmatrix} \mathbf{p}\\ \mathbf{q}\end{pmatrix}= -\mathbf{P}
\begin{pmatrix} \mathbf{c}\\ \mathbf{d}\end{pmatrix},
\quad\mathrm{and}\quad
\begin{pmatrix} \mathbf{a}\\ \mathbf{b}\end{pmatrix}=\mathbf{Q}
\begin{pmatrix} \mathbf{c}\\ \mathbf{d}\end{pmatrix}.
\end{equation}
The $T$-matrix can therefore also be obtained from
\begin{equation}
\mathbf{T}=-\mathbf{P}\mathbf{Q}^{-1}.
\label{EqnTPQ}
\end{equation}
This expression provides the basis for one of the most common approaches to calculating the $T$-matrix in practice, namely
the Extended Boundary Condition Method (EBCM) or Null-Field Method \cite{1971WatermanPRD,1975BarberAO,2002Mishchenko,2000Tsang}.
Within this approach, the matrix elements of $\mathbf{P}$ and $\mathbf{Q}$ are obtained analytically as surface integrals over the
particle surface of VSWF cross-products.

\subsection{Symmetry, unitarity, and energy conservation}

The $T$-matrix satisfies \cite{1971WatermanPRD,2002Mishchenko} symmetry relations related to optical reciprocity,
along with unitarity relations related energy conservation, i.e.~the fact that the extinction cross-section $\sigma_\mathrm{ext}$ is the sum of
scattering $\sigma_\mathrm{sca}$ and absorption $\sigma_\mathrm{abs}$
(note that this is related to the optical theorem \cite{1983Bohren,1976NewtonAJP,1979ChylekAO}).
The optical reciprocity relations are typically easy to check and enforce as they are related (see App.~\ref{AppReciprocity}) 
to ensuring the symmetry of certain matrices \cite{1971WatermanPRD}.
The energy conservation condition is in general more problematic.
It is typically expressed by introducing the $S$-matrix (scattering matrix) defined
as $\mathbf{S}=\mathbf{I}+2\mathbf{T}$.
For lossless (non-absorbing) scatterers (for which $\mathrm{Im}(\epsilon_2)=0$), it can then be shown that energy conservation 
is equivalent to $\mathbf{S}$ being unitary \cite{1971WatermanPRD,2002Mishchenko}.
In terms of the $T$-matrix itself, this results in the somewhat more cumbersome
condition:
\begin{equation}
\mathbf{T}+\mathbf{T}^\dagger=-2\mathbf{T}^\dagger\mathbf{T},
\label{EqnUnitarityT}
\end{equation}
which can be viewed as the matrix form of the generalized optical theorem \cite{1976NewtonAJP,1966Newton}.

In EM scattering, absorbing or conducting scatterers,
for which $\mathrm{Im}(\epsilon_2)>0$ (we exclude the special case of perfect conductors here),
are also often considered and the equality above no longer holds.
In this general case, the inequality $\sigma_\mathrm{ext} \ge \sigma_\mathrm{sca}$ then requires that
$\mathbf{I}-\mathbf{S}^\dagger \mathbf{S}$ be a Hermitian positive semi-definite (HPSD) matrix (note that it is Hermitian by construction) \cite{2002Mishchenko},
which results in a relatively cumbersome condition for $\mathbf{T}$.
The energy conservation conditions for $\mathbf{T}$ can therefore be summarized as:
\begin{equation}
\begin{array}{rl}
\mathrm{\it Lossless:} & \quad \mathbf{T}+\mathbf{T}^\dagger=-2\mathbf{T}^\dagger\mathbf{T},\\[0.2cm] 
\mathrm{\it General:} & ~[-\mathbf{T}-\mathbf{T}^\dagger-2\mathbf{T}^\dagger\mathbf{T}]\mathrm{~~HPSD}.
\end{array}
\label{EqnEnergyT}
\end{equation}

\section{The $K$-matrix}
\label{SecKmatrix}

\subsection{Definition}

We here highlight an alternative formulation of the $T$-matrix method, which simplifies the energy conservation condition
and naturally provides a connection with the radiative correction.
Note that we will not here attempt to give a rigorous mathematical
derivation, but rather focus on the new physical insights.
Our proposed formulation is related to the reactance matrix or $K$-matrix,
which can be formally defined as the Cayley transform of the $S$-matrix \cite{1968LandshoffJMP} and has been previously discussed
in the context of the general quantum theory of scattering \cite{1966Newton,1972Taylor}.
Explicitly, we have:
\begin{align}
{\mathbf{K}}=i(\mathbf{I}-\mathbf{S})(\mathbf{I}+\mathbf{S})^{-1}.
\end{align}
A simple consequence of this definition is that $\mathbf{S}$ unitary is equivalent to $\mathbf{K}$ Hermitian.
In terms of the $T$-matrix, we have:
\begin{equation}
{\mathbf{K}} = -i\mathbf{T}(\mathbf{I}+\mathbf{T})^{-1} = -i(\mathbf{I}+\mathbf{T})^{-1}\mathbf{T}.
\end{equation}
%
%
We note that $\mathbf{T}$ and $\mathbf{K}$ commute and we also have
\begin{equation}
\mathbf{K}+\mathbf{K}{\mathbf{T}}=-i{\mathbf{T}}=\mathbf{K}+{\mathbf{T}}\mathbf{K},
\label{EqnExpForInv}
\end{equation}
which, to pursue the analogy with quantum scattering, may be viewed as the matrix version of Heitler's integral equations \cite{1954Heitler,1966Newton,1972Taylor}.
$\mathbf{T}$ can be obtained from $\mathbf{K}$ using
\begin{equation}
{\mathbf{T}} = i\mathbf{K}(\mathbf{I}-i\mathbf{K})^{-1} = i(\mathbf{I}-i\mathbf{K})^{-1}\mathbf{K},
\label{EqnTfromTbar}
\end{equation}
or from the following property:
\begin{equation}
\mathbf{T}^{-1}=-i\mathbf{K}^{-1} - \mathbf{I}.
\label{EqnExpInv}
\end{equation}

It is important to emphasize that the $K$-matrix and $T$-matrix formulations are fully equivalent from
a formal point of view. However, in practice, since approximations are carried out in computing $\mathbf{K}$
and $\mathbf{T}$ (at the very least, truncation of these infinite matrices), the equivalence is no longer
strictly valid. We will in fact show that the $K$-matrix formulation is then the most appropriate one
in approximate treatments where energy conservation needs to remain strictly enforced.
This will lead us naturally to a generalization of the radiative correction procedure discussed earlier.

\subsection{Energy conservation and the $K$-matrix}

It is interesting to discuss the formal implications of the $K$-matrix formulation for
energy conservation.
For non-absorbing scatterers, the unitarity of $\mathbf{S}$, or Eq.~\ref{EqnUnitarityT} in terms of $\mathbf{T}$,
are equivalent to $\mathbf{K}$ being Hermitian:
$\mathbf{K} = \mathbf{K}^\dagger$.
For a general scatterer, the energy conservation condition (Eq.~\ref{EqnEnergyT}) can be shown to be equivalent to
($i\mathbf{K}^\dagger - i\mathbf{K}$) being a Hermitian positive semi-definite matrix (for details see App.~\ref{AppHPSD}). 
More formally, this condition can be restated as $\mathbf{K}$ being a {\it dissipative} matrix \cite{1974Fan,1975Thompson}.
We can therefore rewrite the condition \ref{EqnEnergyT} in terms of $\mathbf{K}$ as:
\begin{equation}
\begin{array}{rl}
\mathrm{\it Lossless:} & \quad \mathbf{K} = \mathbf{K}^\dagger,\\[0.2cm] 
\mathrm{\it General:} & \quad \mathbf{K}\mathrm{~dissipative~} \left([i\mathbf{K}^\dagger - i\mathbf{K}]~\mathrm{~~HPSD}\right).
\end{array}
\label{EqnEnergyK}
\end{equation}

These are much more natural conditions than the ones obtained for $\mathbf{T}$ (or for $\mathbf{S}$).
We note that ($i\mathbf{K}^\dagger - i\mathbf{K}$) is simply, up to a factor of $i/2$, the skew-Hermitian part of $\mathbf{K}$
and the condition is therefore a generalization of $\mathrm{Im}(K) \ge 0$ to the case where $\mathbf{K}$
is a matrix.
We can therefore naturally identify the skew-Hermitian part of
$\mathbf{K}$ as representing absorption while its Hermitian part is linked to scattering/dispersion.
There is a clear analogy with simpler response functions such as the susceptibility $\chi=\epsilon-1$ of a material or the
polarizability $\alpha$ of a scatterer, for which $\mathrm{Im}(\alpha)$ corresponds to absorption (and is zero for lossless cases)
and is subject to the condition: $\mathrm{Im}(\alpha)\ge 0$.
In fact, the requirement that $\mathbf{K}$ be a dissipative matrix suggests that it is the matrix analogue
of a scalar linear response function \cite{1956TollPR} like the polarizability $\alpha$, and should therefore in addition satisfy
causality and dispersion relations \cite{1967Beltrami} akin to Kramers-Kr\"onig relations.
Such mathematical developments are however outside the scope of this work.
We here only point out that in the case of non-absorbing scatterers, the conditions of optical reciprocity
and energy conservation on $\mathbf{K}$ become closely linked (this is because symmetry and hermiticity
become equivalent for real matrices, see App.~\ref{AppReciprocity} for further details).

\subsection{Relation to expansion coefficients}

We now discuss how the $K$-matrix relates to the field expansion coefficients (in terms of VSWFs).
Recall that the $T$-matrix represents the linear connection (see Eq.~\ref{EqnTcoeff}) between the
field expansion coefficients of the scattered field $\mathbf{E}_\mathrm{sca}$
(in terms of outgoing spherical waves with VSWFs $\mathbf{M}^{(3)}_{\nu}$ and $\mathbf{N}^{(3)}_{\nu}$)
and those of the incident field $\mathbf{E}_\mathrm{inc}$ (in terms of regular waves with VSWFs $\mathbf{M}^{(1)}_{\nu}$ and $\mathbf{N}^{(1)}_{\nu}$).
Multiplying Eq.~\ref{EqnExpForInv} by the vector $\bigl( \begin{smallmatrix} \mathbf{a}\\ \mathbf{b}\end{smallmatrix}\bigr)$, we deduce that:
\begin{equation}
\begin{pmatrix} \mathbf{p}\\ \mathbf{q}\end{pmatrix}=i\mathbf{K}
\begin{pmatrix} \mathbf{a}+\mathbf{p}\\ \mathbf{b}+\mathbf{q}\end{pmatrix}.
\label{EqnTbarCoeff}
\end{equation}
The physical meaning of this expression becomes apparent when we expand the total field outside the particle using the basis
$(\mathbf{M}^{(1)}_{\nu}, \mathbf{N}^{(1)}_{\nu}, \mathbf{M}^{(2)}_{\nu}, \mathbf{N}^{(2)}_{\nu})$,
where the latter two VSWFs use the (irregular) spherical Bessel functions of the second kind
(which are superpositions of outgoing and ingoing spherical waves), in contrast
to the usual spherical Hankel functions of the first kind (which are outgoing spherical waves only).
For this, we simply write $\mathbf{M}^{(3)}_{\nu}=\mathbf{M}^{(1)}_{\nu}+i\mathbf{M}^{(2)}_{\nu}$,
which separate the outgoing spherical wave VSWF into a sum of {\it regular} ($\mathbf{M}^{(1)}$) and {\it irregular} ($\mathbf{M}^{(2)}$) contributions
and obtain:
\begin{align}
\mathbf{E}_\mathrm{out}(\mathbf{r})	&= \mathbf{E}_\mathrm{inc}(\mathbf{r}) + \mathbf{E}_\mathrm{sca}(\mathbf{r}) \nonumber\\
	= \sum_{\nu}&{(a_{\nu}+p_{\nu})\mathbf{M}^{(1)}_{\nu}(k_1\mathbf{r}) + (b_{\nu}+q_{\nu})\mathbf{N}^{(1)}_{\nu}(k_1\mathbf{r})} \nonumber\\
	&  \quad + ip_{\nu} \mathbf{M}^{(2)}_{\nu}(k_1\mathbf{r}) + iq_{\nu}\mathbf{N}^{(2)}_{\nu}(k_1\mathbf{r}).
\label{EqnTbarExpansions}
\end{align}

The coefficients $\bigl( \begin{smallmatrix} \mathbf{a}+\mathbf{p}\\ \mathbf{b}+\mathbf{q}\end{smallmatrix}\bigr)$ in Eq.~
\ref{EqnTbarCoeff} can then be interpreted as the the sum of the incident field and the regular part
of the scattered field; the latter can therefore here be viewed as the regularized self-field, i.e.~the non-diverging
part of the field created by the scatterer at its own position ($r=0$).
The $K$-matrix then represents (up to a factor) the linear connection between the
expansion coefficients of the scattered field with those of 
the total field (incident + self-field). 
It can also be viewed mathematically as the linear connection between the expansions coefficients of the
irregular part of the outside field (i.e.~those of $\mathbf{M}^{(2)}_{\nu}, \mathbf{N}^{(2)}_{\nu}$)
and those of its regular part (i.e.~those of $\mathbf{M}^{(1)}_{\nu}, \mathbf{N}^{(1)}_{\nu}$).
This latter remark can be used to show that the $K$-matrix can
be computed as easily as the $T$-matrix in the most common implementation of the $T$-matrix approach, 
the EBCM \cite{1971WatermanPRD,1975BarberAO}.
Explicitly, we can show that (see App.~\ref{AppEBCM} for details):
\begin{equation}
\mathbf{K}=\mathbf{P}\mathbf{U}^{-1},
\end{equation}
where we have introduced the matrix $\mathbf{U}$ such that $\mathbf{Q}=\mathbf{P}+i\mathbf{U}$, which can be computed as easily as $\mathbf{Q}$ by substituting spherical Hankel functions of the first kind,
$h_n^{(1)}(x)=j_n(x)+iy_n(x)$, by  irregular spherical Bessel functions $y_n(x)$ (note that $j_n(x)$ are the regular spherical Bessel functions).
In fact, $i\mathbf{U}$ can be viewed as the irregular part of $\mathbf{Q}$, while $\mathbf{P}$ is its regular part \cite{2002Mishchenko}.
As a result, within the EBCM approach, $\mathbf{K}$ can be calculated as simply as ${\mathbf{T}}$, if not more simply.

One of the central themes of this work is to argue that the formulation of the EM scattering problem in terms of $\mathbf{K}$
is much more than a mere change of notation and presents in some cases many advantages in terms of both
the practical implementations and the physical interpretations of the method.

\section{Formal derivation of the radiative correction}
\label{SecRadCor}

\subsection{Radiative correction and self-field}

The observations in the last section provide a link with the radiative correction \cite{1998Jackson,1982WokaunPRL},
from the point of view of the self-field or self-reaction. Using again the example of a point polarizable dipole as illustration,
the radiative correction can be interpreted as the effect of the self-field, i.e.~the field $\mathbf{E}_\mathrm{SF}$ created by the scatterer onto itself,
which acts in addition to the incident field $\mathbf{E}_\mathrm{inc}$. The induced dipole is therefore written self-consistenly as $\mathbf{p}=\alpha_0(\mathbf{E}_\mathrm{inc}+\mathbf{E}_\mathrm{SF})$, where $\alpha_0$ again denotes the bare (uncorrected) polarizability.
Since by linearity we have $\mathbf{E}_\mathrm{SF}=G\mathbf{p}$ (where $G$ is the electric Green dyadic \cite{2006Novotny} evaluated at the dipole position,
taken isotropic for simplicity), we obtain
$\mathbf{p}=\alpha^\mathrm{RC} \mathbf{E}_\mathrm{inc}$, where the corrected polarizability $\alpha^\mathrm{RC}$ satisfies:
\begin{equation}
\left(\alpha^\mathrm{RC}\right)^{-1} = \alpha_0^{-1}-G.
\label{EqnAlphaSR}
\end{equation}
Classically, $G$ diverges at the dipole
position (which is why regularization is necessary \cite{1998deVriesRMP}), but its imaginary part is finite and can be computed
to recover Eq.~\ref{EqnRadCor}. The $K$-matrix formulation provides a formal generalization of
this approach. As mentioned earlier, the self-field is represented by the regular part of the scattered field,
i.e. the part of its VSWF expansion including regular VSWFs, $\mathbf{M}^{(1)}_{\nu}$ and $\mathbf{N}^{(1)}_{\nu}$ only.
Eq.~\ref{EqnTbarCoeff} is the generalization of $\mathbf{p}=\alpha_0(\mathbf{E}_\mathrm{inc}+\mathbf{E}_\mathrm{SF})$.
The formulation in terms of $\mathbf{K}$ therefore automatically includes this self-reaction. $i\mathbf{K}$ is analogous
to $G\alpha_0$ and represents the bare response, while $\mathbf{T}$ is analogous to $G\alpha^\mathrm{RC}$ and corresponds to
the self-reaction-corrected response, i.e.~the radiative correction.
This analogy is further reinforced by comparing Eq.~\ref{EqnExpInv} and Eq.~\ref{EqnAlphaSR}.
This crucial point can be further developed to provide a rigorous justification of the empirical radiative
correction to the dipolar polarizability using the $K$-matrix, and by extending this concept to more general cases.

\subsection{Energy conservation in approximate treatments}

In practice, a number of approximations may be made when computing the $T$-matrix; at the very least, truncation is necessary.
The energy conservation conditions (Eq.~\ref{EqnEnergyT}) may no longer be satisfied by the computed $T$-matrix, which is clearly undesirable
(it could for example result in a predicted negative absorption cross-section).
The optical reciprocity can be enforced {\it a posteriori} by appropriate symmetrization, but it is more
difficult to enforce energy conservation.
In contrast, the equivalent conditions on $\mathbf{K}$ (Eq.~\ref{EqnEnergyK}) can more easily be enforced even when
approximations are carried out (they are for example conserved upon truncation of the matrix).

As an illustration, the $\mathbf{T}$ or $\mathbf{K}$ matrices may be approximated by expansions
of their matrix elements \cite{1983Bohren},
for example with respect to the size parameter (the lowest order being akin to the quasi-static or Rayleigh approximation),
or with respect to the refractive index (more precisely, $n-1$) for optically soft particles in the Born or Gans approximation \cite{1983Bohren}.
In such instances, the energy condition on $\mathbf{T}$ may no longer be valid, notably because it mixes linear, e.g.~$\mathbf{T}$, and non-linear terms, e.g.~$\mathbf{T}^\dagger\mathbf{T}$. On the other hand, the energy condition on $\mathbf{K}$ only contains linear terms in $\mathbf{K}$
and can be preserved. For example, the approximated $\mathbf{K}$ matrix will remain Hermitian for non-absorbing scatterers.
We can therefore automatically enforce Eq.~\ref{EqnEnergyT} on $\mathbf{T}$ by deriving 
$\mathbf{T}$ from the approximated $\mathbf{K}$ using Eq.~\ref{EqnTfromTbar} or \ref{EqnExpInv}.

The reformulation of the scattering problem in terms of $\mathbf{K}$ therefore provides a natural method to enforce energy conservation
in approximate treatments within the $T$-matrix framework, and this can for example be applied to the problem of radiative correction in EM scattering.


\subsection{General treatment of the radiative correction to the quasi-static approximations}
\label{SecGenRC}

To illustrate further the procedure for deriving the radiative correction within the $K$-matrix approach, we
focus specifically on the important case of the radiative correction to the electrostatics and magnetostatics (or quasi-static) approximations.
These approximations can be obtained from the general solution by taking
the lowest non-zero order terms of the long-wavelength limit as $k_1 \rightarrow 0$.
We will denote $\mathbf{T}^{(0)}$ and $\mathbf{K}^{(0)}$ the corresponding limit of the $T$- and $K$-matrices.
These can in general be obtained from a direct solution of the electrostatics or magnetostatics problem, which is typically
much easier than the full wave solution.

As explained already, in general, the approximate $T$-matrix $\mathbf{T}^{(0)}$ does not satisfy strictly the $T$-matrix energy conservation condition
(Eq.~\ref{EqnEnergyT}); it only satisfies it {\it approximately} to the  accuracy to which it was calculated,
i.e. in the long-wavelength limit as $k_1 \rightarrow 0$.
In contrast, it is relatively straightforward
to ensure that the approximate $K$-matrix $\mathbf{K}^{(0)}$ satisfies {\it exactly} the $K$-matrix energy conservation condition 
(Eq.~\ref{EqnEnergyK}).
If we therefore compute the approximate $T$-matrix from $\mathbf{K}^{(0)}$, using for example Eq.~\ref{EqnExpInv},
the resulting $T$-matrix will automatically satisfy the energy conservation condition and can be
identified with the radiatively corrected $T$-matrix, $\mathbf{T}^{\mathrm{RC}}$, i.e.:
\begin{equation}
\left(\mathbf{T}^\mathrm{RC}\right)^{-1}=-i\left(\mathbf{K}^{(0)}\right)^{-1} - \mathbf{I}.
\label{EqnRCESA}
\end{equation}

These arguments provide a simple procedure to find the expression for the radiative correction for a given problem:
\begin{itemize}
\item Solve the electrostatics and/or magnetostatics problem and find the corresponding $\mathbf{K}^{(0)}$, which should
satisfy the energy conservation condition for $\mathbf{K}$ (Eq.~\ref{EqnEnergyK}).
\item Apply Eq.~\ref{EqnRCESA} to find the $T$-matrix with radiative correction.
\end{itemize}

We note that for point scatterers, the first step is in fact implicit in the definition of its EM response,
for example $\mathbf{p}=\alpha_0 \mathbf{E}$ for an electric dipole.
We also note that the matrix elements of $\mathbf{T}^{(0)}$ and $\mathbf{K}^{(0)}$ are of order $k_1^3$ or higher. As a result, the product $\mathbf{K}^{(0)}\mathbf{T}^{(0)}$ is of order at least $k_1^6$, and from Eq.~\ref{EqnExpForInv}, we therefore have the following approximation
for the matrix elements to lowest non-zero order:
\begin{equation}
\mathbf{K}^{(0)}\equiv -i\mathbf{T}^{(0)},
\label{EqnK0}
\end{equation}
which can be used in many cases to find $\mathbf{K}^{(0)}$ with standard methods and results in:
\begin{equation}
\left(\mathbf{T}^\mathrm{RC}\right)^{-1}=\left(\mathbf{T}^{(0)}\right)^{-1} - \mathbf{I}.
\label{EqnRCESA2}
\end{equation}

This method can be used to generalize the concept of radiative correction to any
type of scatterer, punctual or bodies of arbitrary shapes, including arbitrary multipole orders and interactions between multipoles.
Specific examples that have been recently studied by other means include
point multipoles (quadrupole, etc.) \cite{2009Colas}, bianisotropic lossless point dipole scatterers \cite{2003Belov} and point magnetic and electric dipoles
with magneto-electric coupling \cite{2011SersicPRB}. The expressions obtained in all these studies are in fact special cases of Eq.~\ref{EqnRCESA}
as we shall explicitly show in the next section.
It is also interesting to note that Eqs.~\ref{EqnExpInv} and \ref{EqnRCESA} can equally apply to study the radiative correction
to higher order expansions of the polarizabilities in terms of $k_1$ (as illustrated in the simple case of spheres in the next section) or to expansions in terms of other parameters (for example $n-1$ for optically soft particles).

\section{Application to specific cases}
\label{SecAppl}

We now study in more detail how the arguments presented so far can be applied to
specific cases of interest, some of which have been studied in the past using mostly heuristic arguments.
As we shall see, all examples of radiative corrections studied so far in the literature
are special cases of Eq.~\ref{EqnExpInv} or \ref{EqnRCESA}. The only difficulty is
to link the $T$-/$K$-matrix formalism to more natural physical representations in terms
of for example polarizability, multipole moments, and multipole fields.
We therefore first show explicitly that this link is
relatively straightforward; in essence it is simply a matter of definition and units.
We then focus in the rest of this section on specific examples.

\subsection{Physical interpretations of the vector spherical wave functions}
\label{SecPhys}

The $T$- and $K$-matrix provide relations between coefficients of the expansions of the fields
in vector spherical wave functions (VSWFs). In practice, however, the excitation and response of the system
are typically expressed in a more natural form. For example, the excitation may be in the form of
a constant external electric field (in electrostatics) or the field of a plane wave.
The response is often modeled in the form of an induced dipole (or multipole), the electromagnetic field
of which implicitly represents the scattered field.
For the applicability of the formalism, it is therefore necessary to link these physical excitations/reponses to
their VSWF expansions.
The VSWFs of the scattered field, $(\mathbf{M}^{(3)}_{nm}$ and $\mathbf{N}^{(3)}_{nm})$, are outgoing spherical
waves and can be readily identified \cite{1998Jackson}
with multipolar fields of order $n$ (total angular momentum) and angular momentum number $m$.
$\mathbf{N}^{(3)}_{nm}$ correspond to electric multipoles (also called transverse-magnetic \cite{1998Jackson}),
while $\mathbf{M}^{(3)}_{nm}$ correspond to magnetic multipoles (also called transverse-electric).
The expansion coefficients of the scattered field ($p_{nm},q_{nm}$) are therefore proportional
to the magnetic and electric multipole moments of the scattered field (in a spherical tensor representation).

In a similar fashion, the expansion of the incident field in terms of a series of regular VSWFs,
$(\mathbf{M}^{(1)}_{nm},\mathbf{N}^{(1)}_{nm})$ can be viewed as the multipolar decomposition of the incident field,
also in a spherical tensor representation. For an arbitrary incident plane wave, such an expansion can be computed
analytically \cite{2002Mishchenko} and it is in fact one of the necessary steps in Mie theory \cite{1983Bohren}.

In the context of the radiative correction to the electrostatics approximation, it is interesting
to write explicitly these expressions.
If we consider a general external electric field (with sources at infinity),
which is defined by an electric potential $\phi_\mathrm{inc}(\mathbf{r})$
solution of Laplace equation, we may expand it as (the negative sign is chosen for convenience):
\begin{equation}
\phi_\mathrm{inc}(\mathbf{r})=-\sum_{n,m} \tilde{b}_{nm} r^n Y_{nm}(\theta,\phi),
\end{equation}
where $Y_{nm}(\theta,\phi)$ are normalized scalar spherical harmonics (see App.~\ref{AppVSWF}).
The electrostatic response of a scatterer (point or body) to this external field can be written as
a standard multipole expansion \cite{1998Jackson} of the potential created outside the scatterer as:
\begin{equation}
\phi_\mathrm{sca}(\mathbf{r})=\frac{1}{4\pi\epsilon_0\epsilon_1} \sum_{n,m} \tilde{q}_{nm} \frac{Y_{nm}(\theta,\phi)}{r^{n+1}} .
\end{equation}
The electric fields can be obtained from the standard relation $\mathbf{E}=-\boldsymbol{\nabla}\phi$. 
In the general case, the induced multipole moments (represented as a vector $\tilde{\mathbf{q}}$) are linearly related to the excitation
coefficients $\tilde{\mathbf{b}}$ by:
\begin{equation}
\tilde{\mathbf{q}}={\boldsymbol{\alpha}}\tilde{\mathbf{b}}.
\end{equation}
$\boldsymbol{\alpha}$ is a generalized multipolar static polarizability tensor in the spherical tensor representation.
We note that different proportionality constants (potentially depending on $n,m$) could be introduced
in the multipole expansions above and would affect the definition of $\boldsymbol{\alpha}$.
In fact, the $T$-matrix formulation in the electrostatics limit ($k_1\rightarrow 0$) is an example
of such an alternative definition.
More explicitly, we can obtain the electrostatics limit for the normalized VSWFs (electric multipoles only) as:
\begin{align}
\tilde{\mathbf{N}}^{(1)}_{nm} &=\frac{k_1^{n-1}}{(2n+1)!!}\sqrt{\frac{n+1}{n}}\boldsymbol{\nabla}(r^n Y_{nm}),\nonumber\\
\tilde{\mathbf{N}}^{(3)}_{nm} &=\frac{i(2n-1)!!}{k_1^{n+1}}\sqrt{\frac{n}{n+1}}\boldsymbol{\nabla}\left(\frac{Y_{nm}}{r^{n+1}}\right).
\end{align}
The electrostatics problem can therefore be recast within the $T$-matrix formulation as
\begin{align}
\mathbf{E}_\mathrm{inc}&=\sum_{n,m}b_{nm} \tilde{\mathbf{N}}^{(1)}_{nm},\nonumber\\
\mathbf{E}_\mathrm{sca}&=\sum_{n,m}q_{nm} \tilde{\mathbf{N}}^{(3)}_{nm},
\end{align}
with
\begin{align}
\mathbf{q}=\tilde{\mathbf{T}}^{22}\mathbf{b},
\end{align}
where $\tilde{\mathbf{T}}^{22}$ is the electrostatics limit of $\mathbf{T}^{22}$ (it is the bottom right block of $\mathbf{T}^{(0)}$).
Note that $\mathbf{T}^{11}$, $\mathbf{T}^{12}$,
and $\mathbf{T}^{21}$ involve magnetic multipoles and are zero in an electrostatics problem.

Combining all this, we obtain a relation
between the matrix elements of the multipolar static polarizability tensor and the electrostatics limit of the $T$-matrix:
\begin{align}
\tilde{T}^{22}_{nm}=\frac{ik^{2n+1}}{(2n-1)!! (2n+1)!!}\frac{n+1}{n}\frac{1}{4\pi\epsilon_0\epsilon_1}\alpha_{nm}.
\label{EqnTAlpha}
\end{align}
The matrix $\tilde{\mathbf{T}}^{22}$ is therefore simply, up to some proportionality factors, the multipolar static polarizability tensor.
The radiative correction to the $T$-matrix (Eq.~\ref{EqnRCESA}) therefore also applies
to any definition of the multipolar polarizability tensors except for the proportionality constants
(and correctly keeping track of these proportionality constants is the primary difficulty in
writing it out explicitly). We will give specific examples in the following.

Finally, a similar result can be obtained for the magnetostatics case in terms of the multipolar magnetic polarizability tensor $\beta_{nm}$,
by substituting $\alpha_{nm} \equiv \epsilon_0\epsilon_1 \beta_{nm}$:
\begin{align}
\tilde{T}^{11}_{nm}=\frac{ik^{2n+1}}{(2n-1)!! (2n+1)!!}\frac{n+1}{n}\frac{1}{4\pi}\beta_{nm}.
\label{EqnT11Alpha}
\end{align}

\subsection{Electric and magnetic point dipoles}

The simplest example of radiative correction is that of a point polarizable dipole with polarizability tensor $\boldsymbol{\alpha}_0$, typically
obtained from a quasi-static (electrostatics) treatment.
The response of such a dipole to an incident field $\mathbf{E}_\mathrm{inc}$ at its position is defined by the induced dipole moment: $\mathbf{p}=\boldsymbol{\alpha}_0\mathbf{E}_\mathrm{inc}$.
The $T$-matrix is here simply proportional to the polarizability tensor as given by Eq.~\ref{EqnTAlpha} with $n=1$:
\begin{equation}
\mathbf{T}^{22}=\frac{ik_1^3 }{6\pi\epsilon_0\epsilon_1}\boldsymbol{\alpha},
\end{equation}
which is valid both for the approximate $T$-matrix $\mathbf{T}^{(0)}$ in terms of $\boldsymbol{\alpha}_0$ and
the corrected $T$-matrix $\mathbf{T}^\mathrm{RC}$ in terms of $\boldsymbol{\alpha}^\mathrm{RC}$.
The other blocks of the $T$-matrix are zero in this case.
Moreover, $\mathbf{T}^{(0)}$ is of order $k_1^3$ and we therefore have $\mathbf{K}^{(0)}=-i\mathbf{T}^{(0)}$ (Eq.~\ref{EqnK0}).
We note that for any physical polarizability tensor, $\boldsymbol{\alpha}_0$ should be Hermitian if there is no absorption
and a dissipative matrix if there is absorption (in a diagonal basis with eigenvalues $\alpha_i$, this is equivalent to $\alpha_i$
real, or $\mathrm{Im}(\alpha_i)>0$, respectively). $\mathbf{K}^{(0)}$ therefore satisfies the energy conservation conditions (Eq.~\ref{EqnEnergyK}).
We can therefore rewrite Eq.~\ref{EqnRCESA} in terms of $\boldsymbol{\alpha}$ to obtain the expression for the radiative
correction as
\begin{equation}
\left(\boldsymbol{\alpha}^\mathrm{RC}\right)^{-1}=\left(\boldsymbol{\alpha}_0\right)^{-1}-i\frac{k_1^3}{6\pi\epsilon_0 \epsilon_1}\mathbf{I},
\label{EqnRadCorAgain}
\end{equation}
which is the same as Eq.~\ref{EqnRadCor} previously obtained heuristically for an isotropic polarizability tensor.
The expression above in fact extends it to the case of a general polarizability tensor.
Moreover, the argument remains valid for a body scatterer, when considering only the electric dipolar response.
This for example justifies the empirical use of such a radiative correction for spheroidal particles \cite{2003KellyJPCB,2009MorozJOSAB}.

Note that in general $\alpha_0$ depends on the frequency $\omega$ and the causality condition \cite{1956TollPR} for $\alpha^\mathrm{RC}$ cannot be easily assessed by inspection of Eq.~\ref{EqnRadCorAgain} only. We speculate that, based on the arguments in Ref. \onlinecite{1967Beltrami}, the fact
that $\mathbf{K}$ and its approximation are dissipative will automatically enforce causality for $\alpha^\mathrm{RC}$, but further work (outside the scope of this paper) is necessary to investigate such aspects.

Finally, for a magnetic point dipole, whose response to an incident magnetic field $\mathbf{H}_\mathrm{inc}$ is an induced
magnetic dipole moment $\mathbf{m}=\boldsymbol{\beta}_0 \mathbf{H}_\mathrm{inc}$, we would obtain following
a similar reasoning:
\begin{equation}
\left(\boldsymbol{\beta}^\mathrm{RC}\right)^{-1}=\left(\boldsymbol{\beta}_0\right)^{-1}-i\frac{k_1^3}{6\pi}\mathbf{I}.
\end{equation}

\subsection{Electric point quadrupole and higher order multipoles}

It is straightforward to extend the arguments above to higher order multipoles.
For electric multipoles, we can use Eq.~\ref{EqnTAlpha}, which relates the multipolar polarizability in the spherical tensor representation
to the $T$-matrix. Applying the general formula \ref{EqnRCESA}, we obtain the radiative correction for
an electric multipole of order $n$ as:
\begin{equation}
\left(\boldsymbol{\alpha}^\mathrm{RC}\right)^{-1}=\left(\boldsymbol{\alpha}_0\right)^{-1}
-\frac{i}{4\pi\epsilon_0\epsilon_1}\frac{k^{2n+1}}{(2n-1)!! (2n+1)!!}\frac{n+1}{n}\mathbf{I},
\end{equation}
which is the same expression as proposed in Ref. \onlinecite{2009Colas} for spheres.
The expression above in fact extends it to the case of a general polarizability tensor (for example anisotropic).

\subsection{Point dipoles with magneto-optic coupling}

Sersic {\it et al.} have recently studied in detail the case of a magnetoelectric point-dipole \cite{2011SersicPRB}
in the context of metamaterials and derived an expression
for the radiative correction (Eq.~(18) in Ref. \onlinecite{2011SersicPRB}) using empirical arguments based on the optical theorem for this system \cite{2003Belov}.
This expression appears immediately similar to our general formula (Eq.~\ref{EqnRCESA}) and
the equivalence between the two can be demonstrated providing the prefactors and units are
accounted for carefully, as we now show explicitly.

Following Ref. \onlinecite{2011SersicPRB}, the response of the scatterer to an incident electromagnetic field ($\mathbf{E}_\mathrm{inc},\mathbf{H}_\mathrm{inc}$)
is defined by the induced electric ($\mathbf{p}$) and magnetic ($\mathbf{m}$) dipole moments obtained from the
most general linear relation:
\begin{equation}
\begin{pmatrix}\mathbf{p}\\ \mathbf{m} \end{pmatrix}= 
\boldsymbol{\alpha}
\begin{pmatrix}\mathbf{E}_\mathrm{inc}\\ \mathbf{H}_\mathrm{inc} \end{pmatrix}=
\begin{pmatrix}
\boldsymbol{\alpha}_{EE} & \boldsymbol{\alpha}_{EH}\\
\boldsymbol{\alpha}_{HE} & \boldsymbol{\alpha}_{HH}
\end{pmatrix}
\begin{pmatrix}\mathbf{E}_\mathrm{inc}\\ \mathbf{H}_\mathrm{inc} \end{pmatrix},
\end{equation}
which is here written in rationalized units as in Ref. \onlinecite{2011SersicPRB}.
$\boldsymbol{\alpha}$ is a $6\times 6$ polarizability tensor, compactly written in block-matrix notation.

To apply our formalism, we rewrite this definition in 
SI units as follows:
\begin{align}
\begin{pmatrix}\mathbf{p}^\mathrm{SI}\\ \mathbf{m}^\mathrm{SI} \end{pmatrix}= 
\boldsymbol{\alpha}^\mathrm{SI}
\begin{pmatrix}\mathbf{E}^\mathrm{SI}_\mathrm{inc}\\ \mathbf{H}^\mathrm{SI}_\mathrm{inc} \end{pmatrix},
\end{align}
with \footnote{Note that there is a factor $\sqrt{\epsilon_0/\epsilon}$ missing in Table I of Ref. \onlinecite{2011SersicPRB}
for $\alpha_{EH}$.} 
\begin{align}
\boldsymbol{\alpha}^\mathrm{SI} &= 4\pi\epsilon_0\epsilon_1\begin{pmatrix}
\boldsymbol{\alpha}_{EE} & Z\boldsymbol{\alpha}_{EH}\\
\frac{1}{\epsilon_0\epsilon_1Z}\boldsymbol{\alpha}_{HE} & \frac{1}{\epsilon_0\epsilon_1} \boldsymbol{\alpha}_{HH}
\end{pmatrix},
\end{align}

where $Z=\sqrt{\mu_0/(\epsilon_0 \epsilon_1)}=1/(\epsilon_0 c \sqrt{\epsilon_1})$ is the impedance
of the embedding medium (with relative dielectric constant $\epsilon_1$).

Using the arguments of Sec.~\ref{SecPhys}, we may express each block-matrix of the polarizability tensor
in terms of a block-matrix (defined in Eq.~\ref{EqTmatBlock}) of the $T$-matrix for $n=1$ (dipole terms).
For example, using Eqs.~\ref{EqnTAlpha} and \ref{EqnT11Alpha} with $n=1$, we obtain:
\begin{align}
\mathbf{T}^{22}_{n=1}&=\frac{ik_1^3}{6\pi\epsilon_0\epsilon_1}\boldsymbol{\alpha}^\mathrm{SI}_{EE}= \frac{2}{3}ik_1^3 \boldsymbol{\alpha}_{EE},\nonumber\\
\mathbf{T}^{11}_{n=1}&=\frac{ik_1^3}{6\pi}\boldsymbol{\alpha}^\mathrm{SI}_{HH}= \frac{2}{3}ik_1^3 \boldsymbol{\alpha}_{HH}.
\end{align}
For magneto-electric coupling, the arguments of Sec.~\ref{SecPhys} can be applied by noticing
that to change from $\mathbf{E}$ to $\mathbf{H}$ one may make the substitution
$\mathbf{b} \rightarrow (-i/Z)\mathbf{a}$ for the incident field and $\mathbf{q} \rightarrow (-i/Z)\mathbf{p}$
for the scattered field. We then get:
\begin{align}
\mathbf{T}^{21}_{n=1}&=\frac{ik_1^3}{6\pi\epsilon_0\epsilon_1}\frac{-i}{Z}\boldsymbol{\alpha}^\mathrm{SI}_{EH} = \frac{2}{3}ik_1^3 (-i\boldsymbol{\alpha}_{EH}),\nonumber\\
\mathbf{T}^{12}_{n=1}&=\frac{ik_1^3}{6\pi}(iZ)\boldsymbol{\alpha}^\mathrm{SI}_{EH} = \frac{2}{3}ik_1^3 (i\boldsymbol{\alpha}_{HE}).
\end{align}
These results can be written in more concise form as:
\begin{equation}
\mathbf{T}=\frac{2}{3}ik_1^3 
\begin{pmatrix}
\boldsymbol{\alpha}_{HH} & i\boldsymbol{\alpha}_{HE}\\
-i\boldsymbol{\alpha}_{EH} & \boldsymbol{\alpha}_{EE}
\end{pmatrix}.
\end{equation}

Following our procedure presented in Sec.~\ref{SecGenRC}, we therefore obtain from Eq.~\ref{EqnRCESA} the radiative correction formula
for such a magneto-electric scatterer as:
\begin{align}
\begin{pmatrix}
\boldsymbol{\alpha}^\mathrm{RC}_{HH} & i\boldsymbol{\alpha}^\mathrm{RC}_{HE}\\[0.2cm]
-i\boldsymbol{\alpha}^\mathrm{RC}_{EH} & \boldsymbol{\alpha}^\mathrm{RC}_{EE}
\end{pmatrix}^{-1}
=
\begin{pmatrix}
\boldsymbol{\alpha}^{0}_{HH} & i\boldsymbol{\alpha}^0_{HE}\\[0.2cm]
-i\boldsymbol{\alpha}^0_{EH} & \boldsymbol{\alpha}^0_{EE}
\end{pmatrix}^{-1}
-\frac{2}{3}ik_1^3\mathbf{I}.
\end{align}
To recast this expression in terms of the original definition of $\boldsymbol{\alpha}$,
it is necessary to change basis. One
may introduce the unitary matrix $\mathbf{W}=\bigl( \begin{smallmatrix} \mathbf{0} & i\mathbf{I} \\ \mathbf{I} & \mathbf{0} \end{smallmatrix}\bigr)$
and by left-multiplying by $\mathbf{W}$ and right-multiplying by $\mathbf{W}^\dagger$, we obtain:
\begin{equation}
\left(\boldsymbol{\alpha}^\mathrm{RC}\right)^{-1}= \left(\boldsymbol{\alpha}_0\right)^{-1}-\frac{2}{3}ik_1^3\mathbf{I},
\end{equation}
which is the same expression as previously obtained empirically (Eq.~(18) in Ref. \onlinecite{2011SersicPRB}).


\subsection{Beyond the electrostatics approximation}

%

As mentioned earlier, our procedure for radiative correction can be applied to other types of approximations,
for example beyond the electrostatic approximation.
As an illustration, we will here study in detail the case of a spherical scatterer, for
which exact results can be obtained from Mie theory \cite{1983Bohren}.
In this case, the $P$-, $Q$-, $U$-, $K$-, and $T$-matrix are diagonal and independent of $m$, and the only
non-zero terms are:
\begin{align}
P^{11}_{nn}&=A_n\left[s\psi_n(x)\psi'_n(sx) -\psi'_n(x)\psi_n(sx)\right],\nonumber\\
Q^{11}_{nn}&=A_n\left[s\xi_n(x)\psi'_n(sx) -\xi'_n(x)\psi_n(sx)\right],\nonumber\\
U^{11}_{nn}&=A_n\left[s\chi_n(x)\psi'_n(sx) -\chi'_n(x)\psi_n(sx)\right],\nonumber\\[0.2cm]
P^{22}_{nn}&=A_n\left[\psi_n(x)\psi'_n(sx) -s\psi'_n(x)\psi_n(sx)\right],\nonumber\\
Q^{22}_{nn}&=A_n\left[\xi_n(x)\psi'_n(sx) -s\xi'_n(x)\psi_n(sx)\right],\nonumber\\
U^{22}_{nn}&=A_n\left[\chi_n(x)\psi'_n(sx) -s\chi'_n(x)\psi_n(sx)\right],\nonumber\\[0.2cm]
T^{ii}_{nn}&=-P^{ii}_{nn}/Q^{ii}_{nn},\nonumber\\
K^{ii}_{nn}&=P^{ii}_{nn}/U^{ii}_{nn},
\end{align}
where 
\begin{align}
A_n=i\frac{n(n+1)}{s},
\end{align}
$x=k_1 a$ (with $a$ the radius of the sphere), and $s=\sqrt{\epsilon_2}/\sqrt{\epsilon_1}$ is the relative
refractive index.
The functions $\psi_n(x)$, $\chi_n(x)$, and $\xi_n(x)$ are the Riccati-Bessel functions \cite{1983Bohren}
defined in terms of the spherical Bessel and Hankel functions as:
\begin{align}
\psi_n(x)=xj_n(x),\qquad\chi_n(x)=xy_n(x),\nonumber\\
\xi_n(x)=xh_n^{(1)}(x)=\psi_n(x)+i\chi_n(x).
\end{align}

This is consistent with standard approaches to Mie theory since
\begin{align}
p_{nm}= \Gamma_n a_{nm},\nonumber\\
q_{nm}= \Delta_n b_{nm},
\end{align}
where $\Gamma_n=T^{11}_{nn}=-P^{11}_{nn}/Q^{11}_{nn}$ and $\Delta_n=T^{22}_{nn}=-P^{22}_{nn}/Q^{22}_{nn}$ are the magnetic and electric Mie susceptibilities:
\begin{align}
\Gamma_n&=-\frac
{s\psi_n(x)\psi'_n(sx) -\psi'_n(x)\psi_n(sx)}{s\xi_n(x)\psi'_n(sx) -\xi'_n(x)\psi_n(sx)},\nonumber\\
\Delta_n&=-\frac
{\psi_n(x)\psi'_n(sx) -s\psi'_n(x)\psi_n(sx)}{\xi_n(x)\psi'_n(sx) -s\xi'_n(x)\psi_n(sx)}.
\end{align}

The extinction, scattering, and absorption coefficients (i.e.~cross-sections normalized to the geometrical cross-section)
can be obtained from these as:
\begin{align}
Q_\mathrm{ext}&=-\frac{2}{x^2}\sum_n(2n+1)\left[\mathrm{Re}(\Gamma_n)+\mathrm{Re}(\Delta_n)\right],\nonumber\\
Q_\mathrm{sca}&=\frac{2}{x^2}\sum_n(2n+1)\left[|\Gamma_n|^2+|\Delta_n|^2\right],\nonumber\\
Q_\mathrm{abs}&=Q_\mathrm{ext}-Q_\mathrm{sca}= -\frac{2}{x^2}\sum_n(2n+1)\nonumber\\
& \times \left[|\Gamma_n|^2\mathrm{Re}(1+\Gamma_n^{-1})+|\Delta_n|^2\mathrm{Re}(1+{\Delta_n^{-1}})\right].
\end{align}
Energy conservation, $Q_\mathrm{ext}=Q_\mathrm{sca}+Q_\mathrm{abs}$ with $Q_\mathrm{abs}\ge 0$, then requires that \cite{1979ChylekAO}:
\begin{equation}
1+\mathrm{Re}( \Delta_n^{-1})\le 0,
\label{EqnMieOT}
\end{equation}
with the equality holding for non-absorbing spheres (for which $s$ is real).
Note that this condition is simply Eq. \ref{EqnEnergyT} in the special case of spherical scatterers.
In terms of the $K$-matrix, it takes a simpler form $\mathrm{Im}(K^{ii}_{nn})\ge 0$, where $K^{ii}_{nn}$
is up to a sign the same as $\Gamma_n$ and $\Delta_n$ upon substitution of $\xi_n(x)$ by $\chi_n(x)$.

There have been many attempts to find suitable small-argument expansions of the Mie susceptibilities \cite{1980WiscombeAO,1982WokaunPRL,1983MeierOL,2003KuwataAPL,2003KellyJPCB},
notably in the context of plasmonics for the study of localized surface plasmon resonances (LSPR) in metallic nanospheres,
where $|s|$ may be relatively large.
The main dipolar LSPR is determined by $\Delta_1$ and its resonant character is evident in the
wavelength-dependence of the far-field properties, which are then given by:
\begin{align}
Q_\mathrm{ext}&\approx -\frac{6}{x^2}\mathrm{Re}(\Delta_1),\nonumber\\
Q_\mathrm{sca}&\approx \frac{6}{x^2}|\Delta_1|^2.
\end{align}

The lowest order approximation to $\Delta_1$ is
\begin{equation}
\Delta_1^{(0)} = \frac{2i}{3}\frac{s^2-1}{s^2+2}x^3,
\label{EqnMieESA}
\end{equation}
which is simply equivalent to the electrostatics approximation \cite{1983Bohren}.

\begin{figure*}
\includegraphics[width=\textwidth]{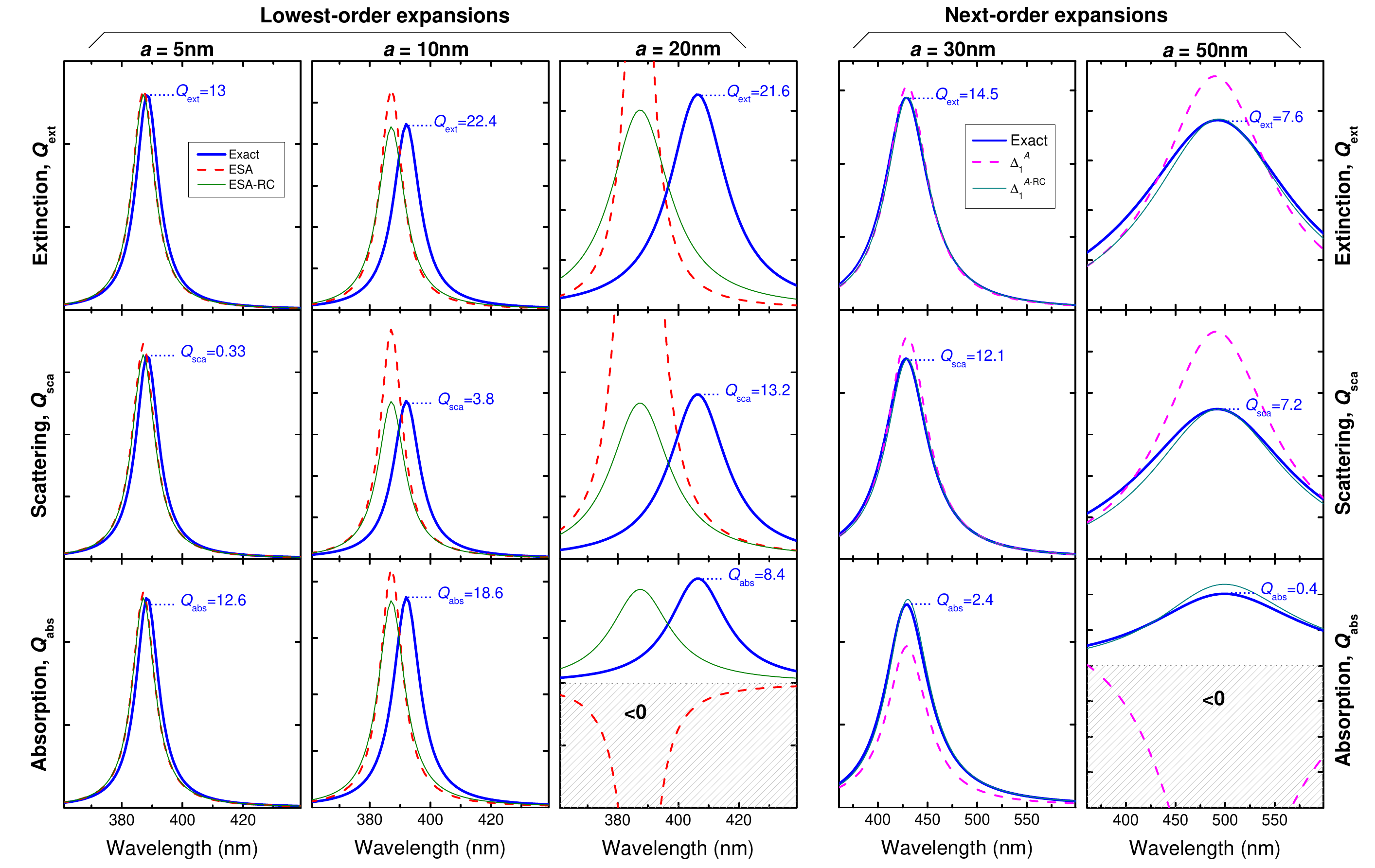}
\caption{(Color online) Predictions of the dipolar localized surface plasmon resonance for a silver nanosphere in water, as evidenced by
the wavelength dependence of the far-field properties: extinction ($Q_\mathrm{ext}$), scattering ($Q_\mathrm{sca}$), and absorption
($Q_\mathrm{abs}=Q_\mathrm{ext}-Q_\mathrm{sca}$). Only the dominant
electric dipole response (corresponding to $\Delta_1$) was included in these calculations. We compare
the exact result (bold/blue lines) with approximate results for increasing sphere size.
For the lowest sizes (radii of $a=5$, 10, and 20\,nm, we compare with the
predictions of the electrostatics approximation (ESA) from Eq.~\ref{EqnMieESA}
(red/dashed lines) and those of the radiative correction to the ESA from Eq.~\ref{EqnMieRCESA} (green/solid lines).
For larger sizes, we compare to the higher order expansions approximations
 using $\Delta_1^{A}$ from Eq.~\ref{EqnMieDelta1A}
(pink/dashed lines) and its proposed radiatively-corrected version $\Delta_1^{A-\mathrm{RC}}$ from Eq.~\ref{EqnMieKDelta1ARC} (dark cyan/solid lines).
Note that these higher order expansions are accurate for $a\le 20$\,nm (their predictions would lie on top of the exact results).
In all cases, the vertical scale has been adjusted for best visualization of the quality of the approximation
and the zero corresponds to the $x$-axis (except in the two cases where the hatched area indicates the negative region).}
\label{FigMieAll}
\end{figure*}

However, as illustrated in Fig.~\ref{FigMieAll} for a silver nanosphere immersed in water, this approximation
is only valid up to very small sizes of $\approx 5$\,nm for metallic spheres, and in fact predicts a negative absorption as $x$ increases.
Moreover, since the electrostatics approximation is size-independent, it does not predict the redshift and broadening
of the LSPR as the size increases. The radiative correction to this dipolar polarizability as given
in Eq.~\ref{EqnRadCor} was in fact originally introduced empirically to remedy this problem \cite{1982WokaunPRL}.
It can be simply expressed as:
\begin{equation}
\left(\Delta_1^{(0)-\mathrm{RC}}\right)^{-1}=\left(\Delta_1^{(0)}\right)^{-1}-1.
\label{EqnMieRCESA}
\end{equation}
and is another example of an application of our general formula \ref{EqnRCESA}.
However, as shown in Fig.~\ref{FigMieAll}, the improvement is marginal and not quantitative. It corrects the problem
of negative absorption as expected, and predicts the strength of the resonance and its broadening, but not the size-induced redshift.

Higher order expansions, up to third relative order, have been proposed, notably \cite{2003KuwataAPL}:
\begin{equation}
\Delta_1^{A}=\Delta_1^{(0)}\frac{1-\frac{x^2}{10}(s^2+1)+O(x^4)}{1-\frac{x^2}{10}\frac{s^2-1}{s^2+2}(s^2+10)
- \Delta_1^{(0)}+O(x^4)}
\label{EqnMieDelta1A}
\end{equation}
In this expression, the numerator and denominator have been expanded to third order (relative to lowest order).
As shown in Fig.~\ref{FigMieAll}, this significantly increases the range of validity of the approximation,
up to $a\approx 20-30$\,nm.
For larger sizes however, although Eq.~\ref{EqnMieDelta1A}
 predicts the correct redshift, it fails to predict the correct magnitude
of the resonance. This can be attributed to the fact
that $\Delta_1^{A}$ does not strictly satisfy the energy conservation condition for the $T$-matrix (Eq.~\ref{EqnEnergyT}),
equivalent to Eq.~\ref{EqnMieOT} here.

The $K$-matrix formalism here provides a simple method to address this issue and improve upon this approximation.
Instead of approximating directly the $T$-matrix ($\Delta_1=-P^{22}_{11}/Q^{22}_{11}$), we therefore use an approximation of the $K$-matrix,
which considering only the electric dipole term is simply $K_1=P^{22}_{11}/U^{22}_{11}$.
Expanding the numerator $P^{22}_{11}$ and denominator $U^{22}_{11}$ as before, we have:
\begin{equation}
K_1^{A}=-i\Delta_1^{(0)}\frac{1-\frac{x^2}{10}(s^2+1)+O(x^4)}{1-\frac{x^2}{10}\frac{s^2-1}{s^2+2}(s^2+10)+O(x^4)}.
\label{EqnMieKDelta1A}
\end{equation}
It is already apparent that the use of the $K$-matrix provides simpler expansions, as the denominator $U^{22}_{11}$ now has a well-defined parity
as opposed to $Q^{22}_{11}$ (this is because $\chi_n(x)$ is odd or even, while $\xi_n(x)$ is not). All odd-order terms in the denominator
therefore disappear (they will in fact reappear as a result of the radiative correction).
$K_1^A$ also satisfies the energy conservation condition, Eq.~\ref{EqnEnergyK}, at least for sufficiently small $x$.
In fact, for $s$ real (non-absorbing sphere), the condition is strictly satisfied for all $x$ since $K_1^A$ is then real.
We apply the central formula of this work, Eq.~\ref{EqnRCESA}, to derive the radiative correction to the approximated $K_1^{A}$ and obtain:
\begin{equation}
\left(\frac{\Delta_1^{A-\mathrm{RC}}}{\Delta_1^{(0)}}\right)^{-1}=\frac{1-\frac{x^2}{10}\frac{s^2-1}{s^2+2}(s^2+10)}{1-\frac{x^2}{10}(s^2+1)}-1.
\label{EqnMieKDelta1ARC}
\end{equation}
This expression differs from the earlier expression (Eq.~\ref{EqnMieDelta1A}) only by terms of relative order $x^4$ or larger.
However, because it was derived from the $K$-matrix formalism, it should be more physically valid, especially
in resonant systems where energy conservation is crucial. This is indeed the case, as it correctly
predicts the LSP resonance behavior better than the previous approach based on a direct expansion of $\Delta_1^{A}$,
in fact up to $a\approx 50$\,nm as shown in Fig.~\ref{FigMieAll}.

\section{Conclusion}

In conclusion, we have shown that the $K$-matrix provides a simple formalism to study general radiative correction problems in
EM scattering. This was demonstrated in this work on two fronts. Firstly, from an abstract point of view, we studied
and highlighted the formal properties of the $K$-matrix, in particular with regard to energy conservation constraints.
And secondly, from a practical point of view, we showed that previously published radiative correction formula
are a straightforward consequence of the $K$-matrix formalism when applying the method described in this work, namely:
rather than obtaining approximations of the $T$-matrix, it is beneficial to derive it from an approximate $K$-matrix
using Eqs.~\ref{EqnExpInv} and \ref{EqnRCESA}.
We expect that other systems will now be able to be studied following the same procedure.
In addition, we believe that the $K$-matrix formulation will play an important role in the general $T$-matrix
approach to EM scattering. For example, as briefly mentioned in the text and in App. \ref{AppEBCM}, it provides an alternative route
to numerical implementations of the $T$-matrix method, which may be
more suited in some situations (for example for non-absorbing scatterers).

\begin{acknowledgments}

The authors are indebted to the Royal Society of New Zealand for support through a Marsden Grant (WRCS and ECLR)
and Rutherford Discovery Fellowship (ECLR).

\end{acknowledgments}

\appendix

\section{VSWF definitions}
\label{AppVSWF}

The four types of VSWFs ($j=1,2,3,4$) are defined using the same convention as Ref. \onlinecite{2002Mishchenko} as follows:
\begin{align}
\mathbf{M}^{(j)}_{nm}&=\gamma_{nm} \boldsymbol{\nabla} \times (\mathbf{r} \psi^{(j)}_{nm}),\nonumber\\
\mathbf{N}^{(j)}_{nm}&=\frac{1}{k_1} \boldsymbol{\nabla} \times \mathbf{M}^{(j)}_{nm}\\
(\mathrm{Note~that:~} &\mathbf{M}^{(j)}_{nm}= \frac{1}{k_1} \boldsymbol{\nabla} \times \mathbf{N}^{(j)}_{nm}),\nonumber
\end{align}
where
\begin{align}
\gamma_{nm}=\sqrt{\frac{(2n+1)\,\,(n-m)!}{4\pi n (n+1)\,\,(n+m)!}}
\end{align}
is a normalization constant
and
\begin{align}
\psi^{(j)}_{nm}(r,\theta,\phi)=z_n^{(j)}(k r)\, P_n^m(\cos(\theta))\,e^{i m\phi}
\end{align}
are solutions of the scalar Helmholtz equation (with wave-vector $k$) in spherical coordinates.
$z_n^{(j)}$ are spherical Bessel functions, the choice of which defines the type of VSWFs (characterized by the superscript $j$):
\begin{itemize}
\item
$z_n^{(1)}=j_n$, i.e.~spherical Bessel function of the first kind, for the regular VSWFs: $j_n(x)\underset{x \rightarrow 0}{=} x^n/(2n+1)!!$ is regular
at the origin.
\item
$z_n^{(2)}=y_n$, i.e.~spherical Bessel function of the second kind, for irregular VSWFs: $y_n(x)\underset{x \rightarrow 0}{\sim} -(2n-1)!!x^{-n-1}$ is irregular
at the origin.
\item
$z_n^{(3)}=h_n^{(1)}=j_n+iy_n$, i.e.~spherical Hankel function of the first kind,
for outgoing spherical wave VSWFs: $h^{(1)}_n(x)\underset{x \rightarrow \infty}{\sim} -(i)^{n+1}e^{ix}/x$.
\item
$z_n^{(4)}=h_n^{(2)}=j_n-iy_n$, i.e.~ spherical Hankel function  of the second kind,
for ingoing spherical wave VSWFs: $h^{(2)}_n(x)\underset{x \rightarrow \infty}{\sim} i^{n+1}e^{-ix}/x$.
\end{itemize}
The associated Legendre functions $P_n^m(\cos(\theta))$ are here defined with the Condon-Shortley phase, i.e.~as:
\begin{equation}
P_n^m(x)=(-1)^m(1-x^2)^{m/2}\frac{\mathrm{d}^m}{\mathrm{d}x^m}P_n(x),
\end{equation}
where $P_n(x)$ are the Legendre polynomials.

Note that in terms of normalized scalar spherical harmonics $Y_{nm}(\theta,\phi)$ \cite{1998Jackson}, we have:
\begin{equation}
\gamma_{nm}\psi^{(j)}_{nm}(r,\theta,\phi)=\frac{1}{\sqrt{n(n+1)}} z_n^{(j)}(k_1 r) Y_{nm}(\theta,\phi).
\end{equation}

%

\section{Energy conservation condition for absorbing scatterers}
\label{AppHPSD}

As mentioned in the main text, for absorbing particles, the inequality $\sigma_\mathrm{ext} \ge \sigma_\mathrm{sca}$ requires that
$\mathbf{I}-\mathbf{S}^\dagger \mathbf{S}$ be a Hermitian positive semi-definite matrix (HPSD) \cite{2002Mishchenko}.
In terms of the $T$-matrix itself, this results in the somewhat cumbersome
condition that the matrix $-\left[\mathbf{T}+\mathbf{T}^\dagger+2\mathbf{T}^\dagger\mathbf{T}\right]$ is HPSD.

A much simpler condition is obtained in terms of the $K$-matrix by noticing that (assuming $\mathbf{K}$ and $\mathbf{T}$ are invertible):
\begin{equation}
\begin{array}{lll}
&\mathbf{I}-\mathbf{S}^\dagger \mathbf{S} & \mathrm{HPSD} \\[0.2cm]
\Leftrightarrow &
\mathbf{K}^\dagger\left[\mathbf{I}-\mathbf{S}^\dagger \mathbf{S}\right] \mathbf{K} & \mathrm{HPSD}\\[0.2cm]
\Leftrightarrow &
-\mathbf{K}^\dagger\left[\mathbf{T}^\dagger (\mathbf{I}+\mathbf{T}) +(\mathbf{I}+\mathbf{T}^\dagger)\mathbf{T}\right] \mathbf{K} & \mathrm{HPSD}\\[0.2cm]
\Leftrightarrow &
-\mathbf{K}^\dagger\mathbf{T}^\dagger (-i\mathbf{T}) -(i\mathbf{T}^\dagger)\mathbf{T} \mathbf{K} & \mathrm{HPSD}\\[0.2cm]
\Leftrightarrow &
\mathbf{T}^\dagger\left[i\mathbf{K}^\dagger -i\mathbf{K}\right]\mathbf{T}  & \mathrm{HPSD}\\[0.2cm]
\Leftrightarrow &
\left[i\mathbf{K}^\dagger -i\mathbf{K}\right] & \mathrm{HPSD}
\end{array}
\end{equation}

The condition for the $K$-matrix is therefore that $\left[i\mathbf{K}^\dagger -i\mathbf{K}\right]$ is Hermitian
positive semi-definite, or equivalently that $K$ is a dissipative matrix \cite{1974Fan}.

Note that the same proof can easily be adapted to prove the special case that $\mathbf{S}$ unitary is equivalent
to $\mathbf{K}$ hermitian.

\section{Computing the $K$-matrix with the Extended Boundary Condition Method (EBCM)}
\label{AppEBCM}

One of the most common approaches to calculating the $T$-matrix in practice is the Extended Boundary Condition Method (EBCM) or Null-Field Method \cite{1971WatermanPRD,1975BarberAO,2002Mishchenko,2000Tsang}.
Within this approach, $\mathbf{T}$ can be conveniently obtained from $\mathbf{T}=-\mathbf{P}\mathbf{Q}^{-1}$ (Eq.~\ref{EqnTPQ}),
where the matrix elements of $\mathbf{P}$ and $\mathbf{Q}$ can be expressed analytically as surface integrals over the
particle surface. Substituting this into Eq.~\ref{EqnExpForInv} and right multiplying by $\mathbf{Q}$, we obtain
$\mathbf{K}(\mathbf{Q}-\mathbf{P})=i\mathbf{P}$.
This leads us to introduce the matrix $\mathbf{U}$ such that $\mathbf{Q}=\mathbf{P}+i\mathbf{U}$ and we then have:
\begin{equation}
\mathbf{K}=\mathbf{P}\mathbf{U}^{-1}.
\label{EqnKPU}
\end{equation}
In addition, the matrix elements of $\mathbf{Q}$ and $\mathbf{P}$ have identical analytical expressions except
for the substitution of $\mathbf{M}_\nu^{(3)}(k_1\mathbf{r})$ for $\mathbf{Q}$ by $\mathbf{M}_\nu^{(1)}(k_1\mathbf{r})$ for $\mathbf{P}$.
The matrix elements of $\mathbf{U}$ can therefore simply be obtained using the same expressions but
now with $\mathbf{M}_\nu^{(2)}(k_1\mathbf{r})$ (this follows from $\mathbf{Q}=\mathbf{P}+i\mathbf{U}$ and 
$\mathbf{M}_\nu^{(3)}(k_1\mathbf{r})=\mathbf{M}_\nu^{(1)}(k_1\mathbf{r})+i\mathbf{M}_\nu^{(2)}(k_1\mathbf{r})$).
Equivalently, $\mathbf{U}$ can be computed like $\mathbf{Q}$, simply substituting
spherical Hankel functions of the first kind, $h_n^{(1)}(x)=j_n(x)+iy_n(x)$, by  
irregular spherical Bessel functions $y_n(x)$.
As a result, within the EBCM approach, $\mathbf{K}$ can be calculated as simply as ${\mathbf{T}}$, if not more simply.

The same conclusion could have been obtained by direct comparison of Eqs.~\ref{EqnTexp}-\ref{EqnTcoeff} for $\mathbf{T}$
with the equivalent expressions \ref{EqnTbarCoeff}-\ref{EqnTbarExpansions}.

In addition, it is interesting to note that the energy conservation condition for lossless scatterers is equivalent within the EBCM to
\begin{equation}
\mathbf{U}^\dagger\mathbf{P}=\mathbf{P}^\dagger\mathbf{U}.
\end{equation}
Defining $\mathbf{Y}=\mathbf{U}^\dagger\mathbf{P}$, this is also equivalent to $\mathbf{Y}$ Hermitian.
In the general case of absorbing scatterers, the condition that $\mathbf{K}$ be a dissipative matrix is also equivalent to
$\mathbf{Y}$ dissipative.
$\mathbf{K}$ can then be obtained from the following expression:
\begin{equation}
\mathbf{K}=\mathbf{P}\mathbf{Y}^{-1}\mathbf{P}^{\dagger},
\end{equation}
which automatically implies that $\mathbf{K}$ is Hermitian (dissipative) if $\mathbf{Y}$ is
Hermitian (dissipative).
These observations allow one to check (and even enforce) the energy conservation condition on $\mathbf{Y}$
(for example as a function of truncation) before carrying out any matrix inversion.
We believe that such an approach may also be further developed to improve the numerical stability of the
$T$-matrix approach, which is a common issue in EBCM implementations \cite{1977Barber,2011SomervilleOL,2012SomervilleJQSRT}.

\section{Equivalence of optical reciprocity and energy conservation conditions on $\mathbf{K}$}
\label{AppReciprocity}

We here restrict ourselves to non-absorbing scatterers.
In order to highlight the central idea without being hampered by technicalities, it is enlightening
to first consider the somewhat artificial case involving electric multipoles only, i.e.~only the block $\mathbf{T}^{22}$ of the 
$T$-matrix, and the case of a scatterer with symmetry of revolution (for which different values of $m$ are decoupled).
In this case, optical reciprocity is equivalent to $\mathbf{K}^{22}$ symmetric
whereas energy conservation is equivalent to $\mathbf{K}^{22}$ Hermitian.
For non-absorbing scatterers, the matrix elements of $\mathbf{P}^{22}$ and $\mathbf{U}^{22}$ are pure imaginary numbers
and therefore $\mathbf{K}^{22}$ is by construction a real matrix (from Eq.~\ref{EqnKPU}). The conditions for optical reciprocity
($\mathbf{K}^{22}$ symmetric) and energy conservation ($\mathbf{K}^{22}$ Hermitian) then
become trivially equivalent.
We note that this equivalence is not obvious when considering the $T$-matrix as opposed to the $K$-matrix.

This argument can in fact be generalized to the full $K$-matrix for a scatterer of arbitrary shape.
The optical reciprocity condition then takes the form:
\begin{equation}
K^{ij}_{n,m,n',m'} = (-1)^{m+m'} K^{ji}_{n',-m',n,-m},
\label{EqnSymK}
\end{equation}
which is deduced from an identical relation for $\mathbf{T}$ (\cite{2002Mishchenko}, Eq.~5.34).

In the framework of the EBCM approach, $\mathbf{K}$ is computed from $\mathbf{K}=\mathbf{P}\mathbf{U}^{-1}$, where
the matrix elements of $\mathbf{P}$ and $\mathbf{U}$ are given by surface integrals involving cross-products of $\mathbf{M}^{(1)}$,
$\mathbf{N}^{(1)}$, $\mathbf{M}^{(2)}$, $\mathbf{M}^{(2)}$, for example:
\begin{equation}
J_{n,m,n',m'}=-i(-1)^m\int_S dS\mathbf{n}\cdot\left[\mathbf{M}^{(1)}_{n',m'}\times\mathbf{M}^{(1)}_{n,-m}\right].
\end{equation}
Moreover, in a non-absorbing medium (with wavevector $k$ real), we have:
\begin{align}
\mathbf{M}^{(1)}_{n,-m}(k\mathbf{r})=(-1)^m \left(\mathbf{M}^{(1)}_{n,m}(k\mathbf{r})\right)^*
\label{EqnMconj}
\end{align}
along with identical relations relating to $\mathbf{N}^{(1)}$, $\mathbf{M}^{(2)}$, and $\mathbf{N}^{(2)}$.
For an integral like the one given above, we therefore have:
\begin{equation}
(J_{n,m,n',m'})^* = - (-1)^{m+m'} J_{n,-m,n',-m'}.
\end{equation}

By inspection of the integrals for the matrix elements of $\mathbf{P}$ and $\mathbf{U}$ (Ref. \cite{2002Mishchenko} p. 145),
we therefore deduce that:
\begin{equation}
(P^{ij}_{n,m,n',m'})^* = -(-1)^{m+m'} P^{ij}_{n,-m,n',-m'},
\end{equation}
and
\begin{equation}
(U^{ij}_{n,m,n',m'})^* = -(-1)^{m+m'} U^{ij}_{n,-m,n',-m'}.
\end{equation}
By carrying out explicitly the block inversion of $\mathbf{U}$ and block-matrix multiplication of $\mathbf{P}\mathbf{U}^{-1}$,
one may then show that
\begin{equation}
(K^{ij}_{n,m,n',m'})^* = (-1)^{m+m'} K^{ij}_{n,-m,n',-m'}.
\end{equation}
Using this expression (only valid for non-absorbing scatterers), it is clear that
the optical reciprocity condition (Eq.~\ref{EqnSymK}) is equivalent to
\begin{equation}
(K^{ij}_{n,m,n',m'})^* = K^{ji}_{n',m',n,m},
\end{equation}
which is exactly the condition for energy conservation, $\mathbf{K}=\mathbf{K}^\dagger$.

\end{document}